\documentclass[12pt]{article}

\usepackage{scicite}
\usepackage{times}
\usepackage{graphicx}

\graphicspath{ {./} }

\newcommand{\Eelastic}{\mathcal{E}_{el}}

\newcommand{\Ebending}{\mathcal{E}_b}
\newcommand{\Einit}{\mathcal{E}_0}
\newcommand{\anl}{\mathcal{A}}

\usepackage{color}

\title{Tensional twist-folding of sheets into multilayered scrolled yarns} 

\author{Julien Chopin$^{1,2,\ast}$ and Arshad Kudrolli$^{1, \ast}$\\
\\
\normalsize{$^{1}$Department of Physics, Clark University, Worcester, MA 01610, USA,}\\
\normalsize{$^{2}$Instituto de F\'isica,  Universidade Federal da Bahia, Salvador-BA 40170-115, Brazil}\\
\\
\normalsize{$^{\ast}$To whom correspondence should be addressed;}
\\
\normalsize{E-mail: julien.chopin@ufba.br (JC); akudrolli@clarku.edu (AK)}
}

\date{}

\begin{document} 


\baselineskip24pt


\section*{Title}
Tensional twist-folding of sheets into multilayered scrolled yarns

\section*{Short title}
Tensional twist-folding and scrolling of sheets


\section*{Authors and affiliations}
Julien Chopin$^{1,2,\ast}$ and Arshad Kudrolli$^{1, \ast}$

\noindent $^{1}$Department of Physics, Clark University, Worcester, MA 01610, USA,\\
$^{2}$Instituto de F\'isica,  Universidade Federal da Bahia, Salvador-BA 40170-115, Brazil\\
$^{\ast}$To whom correspondence should be addressed;
E-mail: julien.chopin@ufba.br (JC); akudrolli@clarku.edu (AK)


\section*{Abstract}
Twisting sheets as a strategy to form functional yarns relies on millennia of human practice in making catguts and fabric wearables, but still lacks overarching principles to guide their intricate architectures. We show that twisted hyperelastic sheets form multilayered self-scrolled yarns, through recursive folding and twist localization, that can be reconfigured and redeployed. We combine weakly nonlinear elasticity and origami to explain the observed ordered progression beyond the realm of perturbative models. Incorporating dominant stretching modes with folding kinematics, we explain the measured torque and energetics originating from geometric nonlinearities due to large displacements. Complementarily, we show that the resulting structures can be algorithmically generated using Schl\"afli symbols for star-shaped polygons. A geometric model is then introduced to explain the formation and structure of self-scrolled yarns. Our tensional twist-folding framework shows that origami can be harnessed to understand the transformation of stretchable sheets into self-assembled architectures with a simple twist.

\section*{Teaser}
{An elasto-geometric twist-folding framework is advanced to make intricate yarns guided by 3D imaging and origami kinematics.}


\section*{Introduction}
\begin{figure*}
	\centering
\includegraphics[width=1\textwidth]{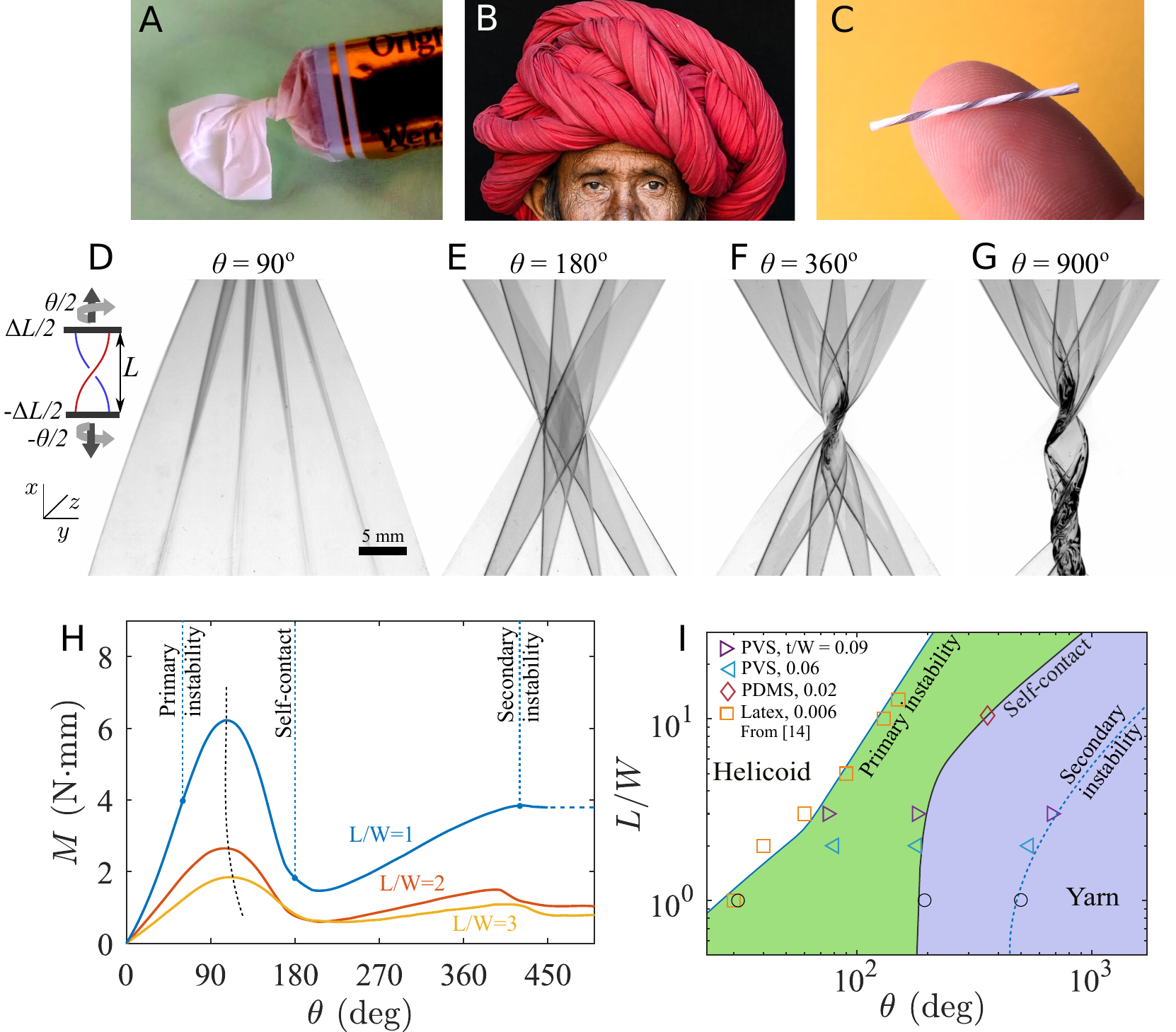}
	\caption{{\bf Experiments reveal a highly ordered transformation to yarns when sheets held under tension are twisted beyond the onset of primary instabilities.} { Examples of twisted, folded, and scrolled structures: ({\bf A}) wrapped candy, ({\bf B})  multifunctional Rajashtani Turban (Photo credit: Lauren Cohen), ({\bf C}) scrolled yarn from a polyethylene sheet (see Supplementary Materials 4: Yarn fabrication from plastic bag).} ({\bf D}-{\bf G}) Shadowgraphs of a transparent PDMS sheet twisted through angle $\theta$ as shown in the inset ($L/W = 1$; $t/W = 0.0028$; $\Delta L/L =  0.1$; $\theta_p = 60 \pm 5^\circ$). Inset: Schematic and lab coordinate system. ({\bf D}) Wrinkles observed just above the onset of primary instability. ({\bf E}) Accordion folded sheet with self-contact. ({\bf F}) A nested helicoid with folded layers develop as the sheet is twisted further. ({\bf G}) Secondary buckling instability occurs with further twisting resulting in a yarn-like structure. The scale bar is the same in ({\bf D-G}). ({\bf H}) The measured torque shows a { repeated increasing and decreasing} sawtooth variation with twist. The amplitude of variation increases as $L/W$ decreases. ({\bf I}) A map delineating regions where the primary instability, self-contact, and secondary instability occur as a function of aspect ratio and twist. Lines are a guide for eyes, except the primary instability for $L/W>3$, which is from Ref.~\cite{Kudrolli2018}.}
\label{fig:intro}
\end{figure*}

Tensional twist-folding is a method to transform flat sheets into layered structures and yarns with ordered internal architectures by remote boundary manipulation. Twisting sheets under tension has been used since antiquity in making catgut bow strings, surgical sutures, musical chord instruments, sports rackets, sausage and candy wrappers, fabric filters and wearbles such as turbans and crushed dupattas, and in the upcycling of plastic (Fig.~\ref{fig:intro}A-C). 
Scrolled yarns with nested structures optimized for  novel applications are being developed as in energy harnessing, batteries, and in embedding  amorphous materials~\cite{kwon2014high,Lima2011,Cruz2014,xu2013ultrastrong}. Such structures are difficult to achieve by compression-induced transformations of elastic sheets without further direct manipulation~\cite{lobkovsky1995scaling,pocivavsek2008stress,silverberg2014using}, and  traditional fiber spinning methods~\cite{hearle1969structural}.  When appropriate materials are used, the transformations can be reversible, and twist folding and scrolling can be used to reconfigure and repurpose flat sheets, as exemplified by the multifunctional Rajasthani turban. 

The interplay between topology and large shape transformations have been studied in terms of inextensible twisted rods and ribbons to understand the conformation of DNA and proteins~\cite{fuller1971writhing,vanderheijden2003instability,kit2012twisting}, and has contributed to the development of a now well-established theoretical framework~\cite{charles2019topology,patil2020topological}. However, shape transformation of sheets
which significantly stretch upon twist have remained undocumented despite their ubiquity in a wide range of applications. 

We report the spontaneous formation of twisted multi-layered yarns with ordered internal architectures enabled by x-ray 3D scanning. These structures obtained under extreme deformation and self-contact are distinct from those observed in rods and ribbons at moderate twist reported previously~\cite{Green1937,Chopin2013}, and are not known to occur by purely compression-driven transformations of elastic sheets as in crumpling, folding, and capillary  wrapping~\cite{Witten2007,Audoly2002,py2007capillary,paulsen2015optimal}.
Modeling such large shape transformations and configurations is extremely challenging.  Elastic plate models such as the F\"oppl-von K\'arm\'an (FvK) equation, and its more recent co-variant extension~\cite{Chopin2015}, have solved the initial growth of wrinkling above onset of primary instability~\cite{davidovitch2011prototypical,vandeparre2011wrinkling,Chopin2015,vella2015indentation,Chopin2016,Kudrolli2018,Panaitescu2018c}, but fail to anticipate, let alone explain, the proposed transformation of a flat sheet into scrolled yarns with functional guests~\cite{Lima2011}. As in other paradigm pattern formation systems such as buoyancy-driven Rayleigh-B\'enard convection which displays intermittent spatio-temporal chaos~\cite{Cross1993}, it is not a priori obvious what imprints of the primary instabilities persist, as a sheet is twisted far beyond the perturbative regime where previous studies were focused. While origami and inextensible sheet models are amenable to address large shape transformations~\cite{BenAmar1997,Cerda1999,hamm2004dynamics,Blair2005,hunt2005twist,py2007capillary,Korte2010,diamant2011compression,Santangelo2017,Callens2018,kumar2018wrapping,paulsen2019wrapping}, their generalization to significantly stretched sheets is unknown.

Going beyond reporting the discovery of ordered transformations, we develop a framework which combines the kinematics of stretched sheets, origami, and fold-induced transverse stiffness to explain our observations. Remarkably, we find that the observed accordion folded sheets have regular polygonal shapes described by Schl\"afli symbols~\cite{coxeter1972regular}, and show that origami kinematics can capture the main features of the structure. We provide an analytical framework to address the successive transformations experienced by a twisted sheet from the onset of transverse wrinkling via recursive folding and  scrolling. Our framework can serve as a guide for the fabrication of yarns with precise control of crosssectional architecture. When made with hyperelastic materials, which recover their unstressed states, they can be repeatedly reconfigured and redeployed with our twist-folding method.

\section*{Results}

\subsection*{Ordered shape-transformation and nonmonotonic torque with twist}
Examples of a polydimethylsiloxane (PDMS) sheet with increasing twist are shown in Fig.~\ref{fig:intro}D-G and Movie~S1. { A system schematic and the Cartesian coordinates system ($x,y,z$) are shown in Fig.~\ref{fig:intro}D, inset.} The system consists of a sheet of length $L$, width $W$, and thickness $t$, twisted by an angle $\theta$ while being held at opposite ends and stretched axially by $\Delta L$.  Transverse wrinkles can be observed just above the onset of primary instability (Fig.~\ref{fig:intro}D), which grow in amplitude and collapse into an accordion folded spiral structure with self-contact (Fig.~\ref{fig:intro}E). As the applied twist is increased further, a nested helical structure forms at the waist (Fig.~\ref{fig:intro}F), and then a secondary instability occurs, which leads to recursive folding and a scrolled multilayered yarn (Fig.~\ref{fig:intro}G). Each of the major shape transformations causes the rate of change of applied torque $M$ to change sign, leading to a sawtooth variation with twist (Fig.~\ref{fig:intro}H). The primary instability and parameter space over which these transformations occur varies with $L/W$ (Fig.~\ref{fig:intro}I). The observed angle at which the primary instability occurs in Fig.~\ref{fig:intro}I is consistent with the bendable ribbon regimes ($L > W$), which scales  as $\theta_p \sim (L/W)^{1/2}$ and wavelength $\lambda_p = \alpha_{\lambda} {\sqrt{Lt}}(\Delta L/L)^{-1/4}$, with $\alpha_{\lambda} = 2.2$ reported previously~\cite{Chopin2015,Kudrolli2018}. The ordered nonperturbative sheet transformations  as the sheets self-fold and form scrolled yarns, the sawtooth torque variation, and the phase-diagram reported in Fig.~\ref{fig:intro}E-I are all documented here for the first time, and are the focus of the analysis to follow.

\subsection*{Tensional twist-folding framework}
\begin{figure*}
	\centering
\includegraphics[width=.9\textwidth]{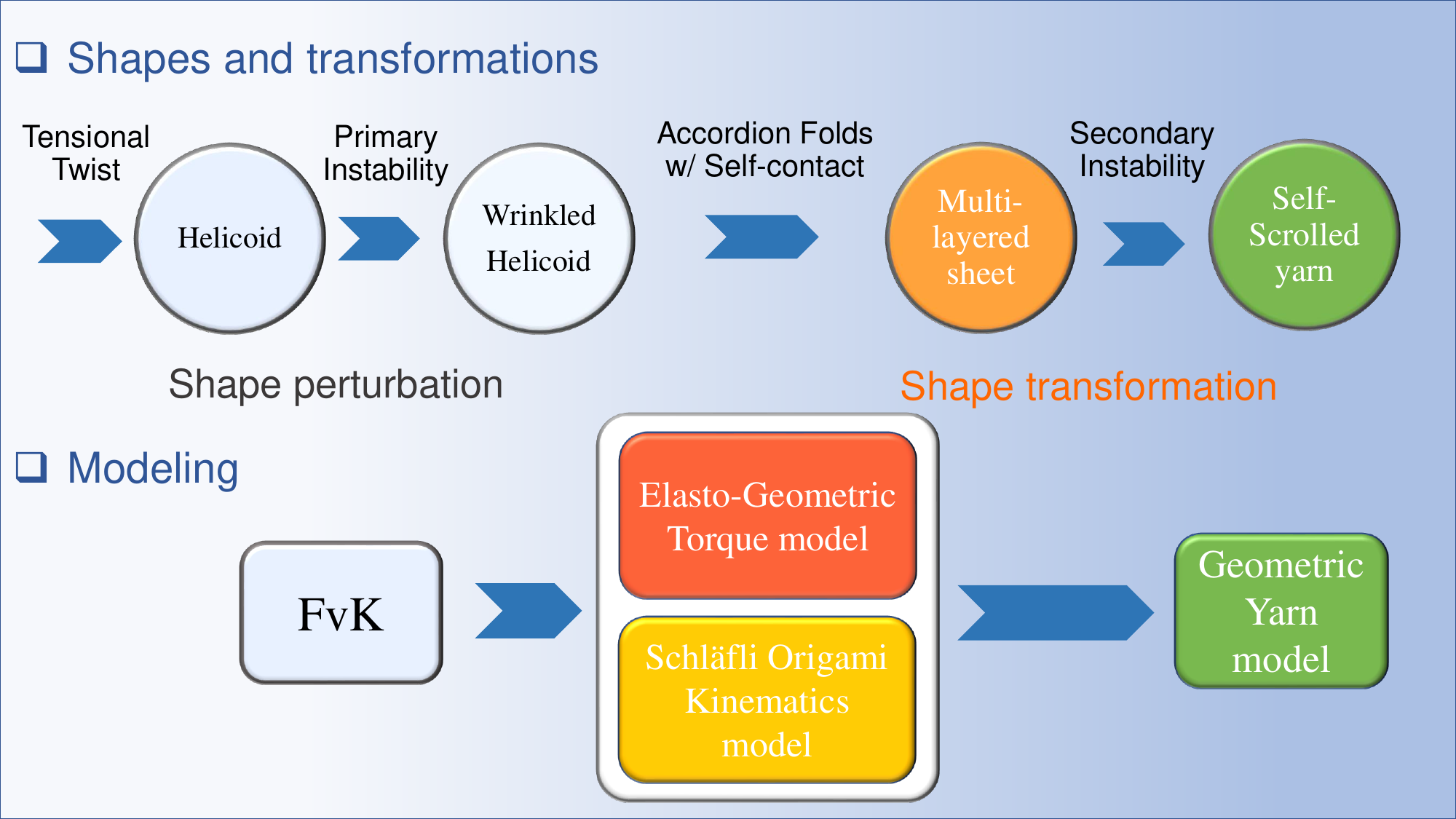}
	\caption{
{\bf {An overview of the observations transformations with twist and the tensional twist folding framework.}}  The observed main transformations as a planar sheet experiences tensional twist-folding and scrolling with applied twist. The elasto-geometric framework including the perturbative FvK formalism, the elasto-geometric torque model which incorporates geometric nonlinearities to explain the stress-strain relation with twist, the Schl\"afli origami kinematic model, and the geometric yarn model.}
	\label{fig:overview}
\end{figure*}

Figure~\ref{fig:overview} gives an overview of the tensional twist-folding framework we introduce to understand the observed main stages of the transformation of a planar sheet into self-scrolled yarns.  The covariant form of the FvK equations, which was introduced recently~\cite{Chopin2015}, can only address longitudinal and transverse wrinkles observed at the onset of primary instabilities~\cite{Green1937,Chopin2013}. Therefore, we introduce a set of models combining geometry, elasticity, and kinematics to capture the observed shapes transformations well beyond perturbations about the helicoidal base state. Building on tensional field theory~\cite{Mansfield1969}, we develop an 
elasto-geometric torque model to capture the stored elastic energy and torsional response beyond the primary instability till the self-contact regime. We then introduce the Schl\"afli origami kinematics model to describe the accordion folded spiral structure that form. Finally, we discuss the geometric yarn model to describe the development of the scrolled yarn structures that form  via helical wrapping with continued multiple twists after secondary instabilities occur.

\subsection*{Curvature localization and accordion folds}
\begin{figure*}
	\centering
\includegraphics[width=.99\textwidth]{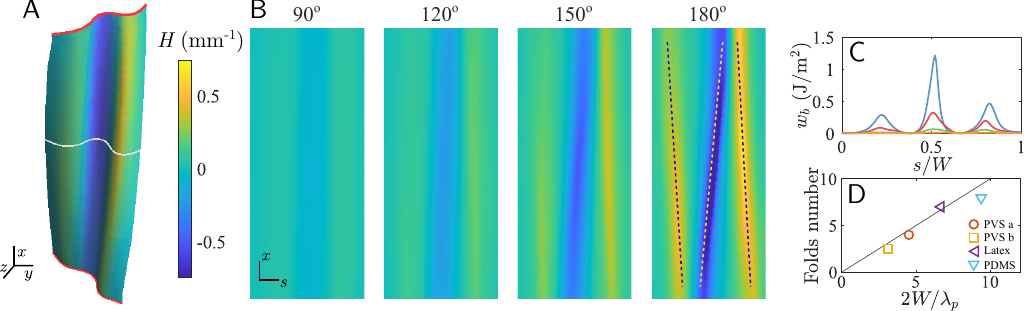}
	\caption{
{\bf {Accordion folding through curvature localization.}} ({\bf A}) The deformation of a PVS sheet twisted by $\theta = 120^\circ$ obtained with x-ray tomography and rendered with mean curvature $H$ given by color bar on right ($L/W = 3$; $t/W = 0.009$; $\theta_p = 75^\circ \pm 5^\circ$). The central 80\% of the sheet away from the clamps is shown.  ({\bf B}) The spatial distribution $H$ mapped to a rectangular domain shows symmetry breaking and localization of the sheet curvature with twist. ({\bf C}) Bending content $w_b$ shows the localization of energy with creasing across the crosssection indicated by the solid white line in (A). ({\bf D}) The measured number of folds $n$ compared with the relation given by the wavelength of the primary instability $n = 2W/\lambda_p$. The aspect ratios $(t/W,L/W)$ are: PVS-a $(0.009,2)$; PVS-b $(0.006,3)$; PDMS $(0.003,1)$; Latex $(0.003,2)$. The three materials are hyperelastic with Young's modulus $E = 1.2$\,MPa (PVS), $6.2$\,MPa (PDMS), and $3.6$\,MPa (Latex).}
	\label{fig:geom}
\end{figure*}

We first discuss in more detail the morphologies of the sheet, obtained with noninvasive 3D x-ray tomography, beyond incipient wrinkling until self-contact. In Fig.~\ref{fig:geom}A, we reconstruct the central 80\% of a twisted poly-vinyl siloxane (PVS) sheet, and render the surface with its mean curvature $H$ (see Supplementary Materials 6: Curvature calculation). High-curvature regions along wrinkle antinodes develop above the onset of transverse instability. To recognize the spatial distribution of the curvature as the sheet wraps around itself, we map $H$ to a rectangular domain with axes $(s,\,x)$, the curvilinear and longitudinal coordinates, respectively (Fig.~(\ref{fig:geom}B), Fig.~S2 and Fig.~S5A).  The wrinkles are observed to be initially aligned with the applied tension when $\theta = 90^\circ$ consistent with linear perturbation analysis~\cite{Chopin2015}.  With increasing twist, $H$ is increasingly localized along folds with essentially flat regions in between, and the folds rotate away from the tensional axis till they meet near the clamped edges. 
 
We calculate the bending energy density  $w_b = B/2 \ ( H^2 + 2(1-\nu)K)$, where $B = E \ t^3/[12(1-\nu^2)]$ is the bending stiffness, $\nu$ the Poisson ratio, $K$ the Gaussian curvature, and $E$ the Young's modulus. Plotting $w_b$ across the sheet at mid-distance between the clamps, we observe sharp peaks grow with $\theta$ showing that the bending energy increasingly localizes along the folds (Fig.~\ref{fig:geom}C). Since the number of peaks is unchanged as twist is increased, we presume that the number of folds $n$ are set by twice the ratio of $W$ and $\lambda_p$. Using sheets with various Young's modulus and aspect ratios, we plot the measured number of folds versus $2W/\lambda_p$ in Fig.~\ref{fig:geom}D. The data collapses onto a line with slope 1, supporting our presumption that the primary instability determines the main features of the shape transformation far above the perturbative wrinkling regime.

\begin{figure}
	\centering
\includegraphics[width=0.95\textwidth]{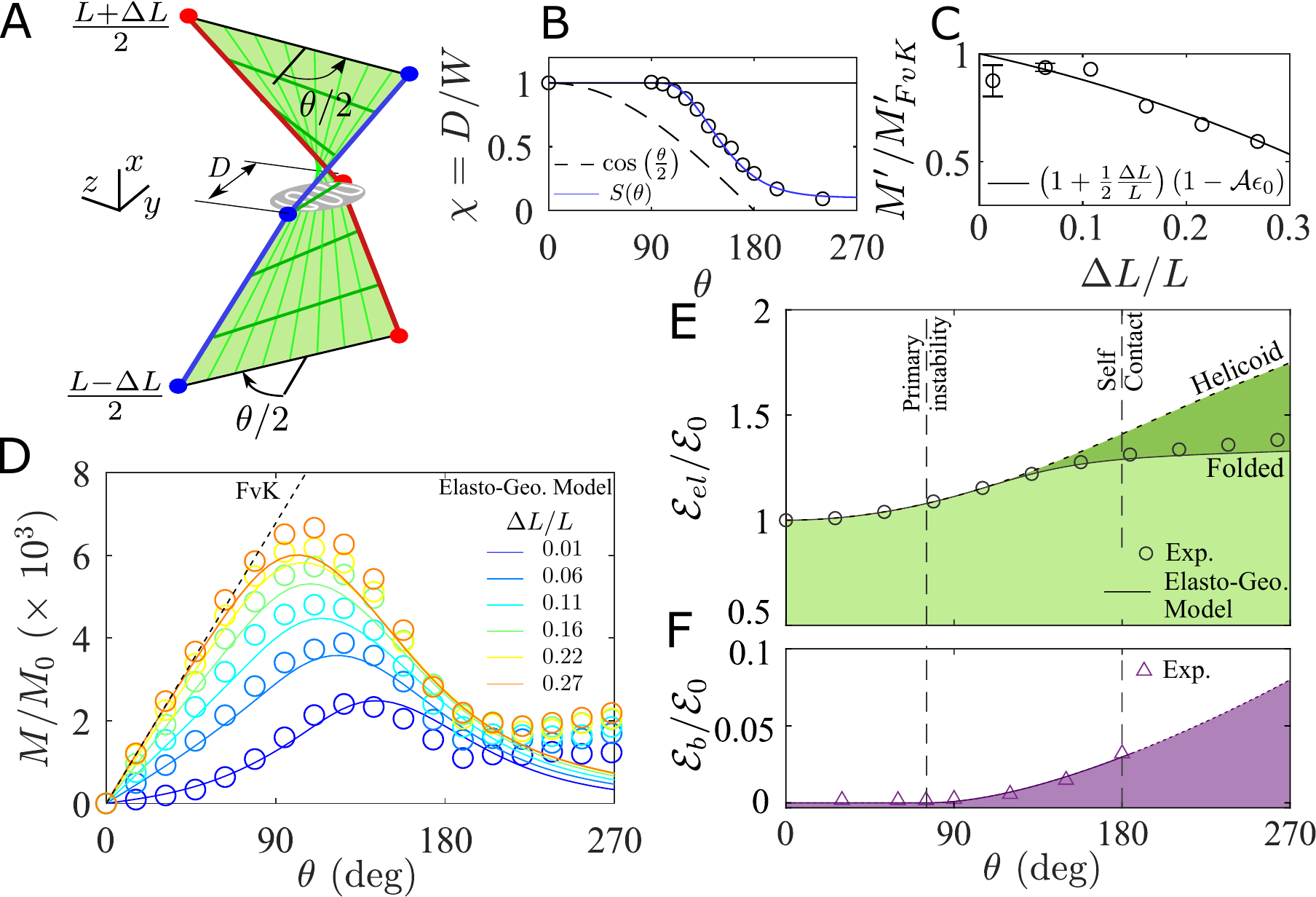}
	\caption{
 {\bf  Elasto-geometric torque model, and comparison with measured torque and energy.}
({\bf A}) Schematics illustrating the backbone structure used in the elasto-geometric model. ({\bf B}) Evolution of the compaction parameter with twist superposed on the sigmoid function $S(\theta)$ defined in Eq.~(S4). ({\bf C}) The measured torsional stiffness $M'$ scaled by the FvK model torsional stiffness $M'_{FvK}$ versus applied strain grows nonlinearly and is described by the strain-softening response of cross-linked elastomers to moderate strain captured by the hyperelastic parameter $\mathcal{A}$. Error bars smaller than symbol size are not shown for clarity. ({\bf D}) The increase and decrease of the measured torque $M$ (circles) versus $\theta$ is quantitatively captured by our elasto-geometric model (solid line). ({\bf E}) Elastic energy  $\Eelastic$ obtained experimentally (circles) and from folded model and unbuckled helicoid scaled by $\Einit$ at zero twist. Elastic energy for $\chi=1$ (helicoid case) is significantly higher for $\theta>180^\circ$ (dashed line). ({\bf F}) The scaled bending energy $\Ebending/\Einit$ (purple triangles) obtained by integrating the measured bending energy density is at least an order of magnitude lower than $\Eelastic/\Einit$. { Solid and dashed lines are guides to the eye.}   }
	\label{fig:model}
\end{figure}

To characterize the transformation with twist, we consider the projected length $D$ onto the $y$-axis of the mid-crossection located mid-way between the two clamps as sketched in Fig.~\ref{fig:model}A. Introducing a dimensionless compaction parameter $\chi = D/W$, we plot the measured $\chi$ as a function of $\theta$ in Fig.~\ref{fig:model}B. We find that $\chi = 1$ as the sheet deforms into a helicoid with diameter $W$, and then starts to decrease for $\theta > \theta_p$ as the sheet undergoes a transverse instability and begins to fold. In the limit of a very thin sheet, i.e. $t \rightarrow 0$, we expect $n \propto W/\lambda_p \rightarrow \infty$, since $\lambda_p \propto \sqrt{t} \rightarrow 0$, and thus $\chi \propto  n t/W \rightarrow 0$ as the sheet folds to the point $(x,y,z)=(0,0,0)$. But for finite $t$, one may expect $\chi$ to decrease and plateau at a finite value as a result of competition between the fold-induced compression resistance, and a stretched-induced compression that determines  $\lambda_p$~\cite{Chopin2015,Kudrolli2018}. 

\subsection*{Elasto-geometric torque model}
Based on these observations, we develop an elasto-geometric model to calculate the stored elastic energy and torsional response of the sheet as it folds beyond the wrinkling regime up until self-contact.  Since the FvK equations are not valid in the large deformation regime, we draw inspiration from tensional field theory~\cite{Mansfield1969,steigmann1990tension} used to describe highly wrinkled sheets where flexural and compressive stresses are negligible compared with tensile stresses. As discussed in detail in the following, we approximate the stretched sheet as it accordion folds with a skeleton surface which enables us to evaluate an analytical expression for the elastic energy $\Eelastic$. We parametrize $\Eelastic$ by defining a compaction parameter $\chi = D/W$, which captures the folding state of the sheet empirically. Then, we obtain the torque $M$ as a function of applied twist using the derivative of the elastic energy with respect to applied twist angle $\theta$ and the measured nonlinear elastic stress-strain constitutive relation.

Guided by the symmetry of the $\pm \theta/2$ twist about the $x$-axis, we consider the end points of the projected mid-crosssection on the $y$-axis ($0, \pm \frac{D}{2}, 0$) used to define $\chi$. Then, joining these two points with the extremities of the upper and lower clamps, as shown by the blue and red lines in Fig.~(\ref{fig:model}A), we obtain a simplified parametrization of the deformed sheet boundary which retains the progressive rotation of the crosssection and its transverse compaction mid-way between the two clamps.

As in tensional field theory~\cite{steigmann1990tension}, we assume that the energetics during folding are predominantly given by the stretching modes in the longitudinal direction while the bending modes are assumed subdominant. Thus, the sheet crosssections are approximated by straight lines with a maximum length $W$ at the clamps and a minimum length $D$ half way between the clamps. Then, the backbone surface - shaded green - is defined as a piecewise linear interpolation between the clamps and the projected mid-crosssection whose parametrization is given by:
\begin{equation}
\vec{r}(x,y) = \frac{|x|}{L/2}\ \vec{r}_c^{\,\,\pm}(y) + \left(1-\frac{|x|}{L/2}\right)\ \vec{r}_{\chi}(y),
\label{eq:r}
\end{equation}
where $\vec{r}_{c}^{\,\,+}$ and $\vec{r}_{c}^{\,\,-}$ parametrizes the upper and lower clamps, respectively, and is given by 
$$\vec{r}_{c}^{\,\,\pm} \equiv \left(\pm\, \frac{L+\Delta L}{2},\,y\cos\left(\frac{\theta}{2}\right),\,\pm\, y\sin\left(\frac{\theta}{2}\right)\right).$$
The mid-crosssection $\vec{r}_{\chi}(y)$ at $x=0$ is parametrized by:
\begin{equation}
\vec{r}_{\chi} = \left(0,\chi(\theta)y,0 \right).
\label{eq:r_chi}
\end{equation}
Eq.~(\ref{eq:r}) can also be interpreted as the parametrization of a piecewise ruled surface where two surfaces are stitched together along the straight mid-crosssection.

In general, the evolution of $\chi$ with twist is unknown for sheets with finite thickness, but we can make an 
estimate in the limit of very thin sheets. The compaction of the sheet is driven by a transverse compressive force, which is expected to vanish with the edge curvature as in the case of a stretched helicoid~\cite{Chopin2013,Chopin2015}. However, since there is no fold-induced resistance to balance compression for very thin sheets, compression vanishes, and the sheet longitudinal edges are essentially straight. In this limit, the mid-crosssection parametrization is simply given as the mean between the top and bottom clamp parametrizations, i.e. $\vec{r}_{\chi} = \left(0, y\cos\left(\frac{\theta}{2}\right), 0\right)$. Hence, $\chi(\theta) = \cos(\frac{\theta}{2})$. We compare this estimate of $\chi$ for very thin sheet with experimental data shown in Fig.~(\ref{fig:model}B), and find that the measured compaction parameter is systematically larger, illustrating the fold-induced resistance to compaction for finite thickness sheets. To account for this effect, we adjusted by the eye a sigmoid function $S(\theta)$ to the measured $\chi$ and use it as an empirical input for our torque model (see Fig.~(\ref{fig:model}B) and Eq.~(S4)).

{
Building on this parametrization of the shape transformation with twist, we calculate the longitudinal Green-Lagrange strain defined as} $\epsilon =\frac{\partial u_x}{\partial x} + \frac{1}{2}\left[\left(\frac{\partial u_x}{\partial x}\right)^2+ \left(\frac{\partial u_y}{\partial x}\right)^2+ \left(\frac{\partial u_z}{\partial x}\right)^2\right]$, where $\vec{u} = (u_x,u_y,u_z)$ is the displacement vector given by $\vec{u}= \vec{r} - (x,y,z)$~\cite{landau}. Using the kinematics given in Eq.~(\ref{eq:r}), the strain reads (see Supplementary Materials 7: Strain): 
\begin{equation}
\epsilon =\epsilon_0+\frac{1}{2}\mathcal{F}^2 \left(\frac{y}{W}\right)^2,
\label{eq:epsilon}
\end{equation}
 where,
$\epsilon_0 = \frac{\Delta L}{L}  + \frac{1}{2}\left(\frac{\Delta L}{L}\right)^2$
is the $\theta$-independent nonlinear strain associated with displacements along the $x$-axis, and
\begin{equation}
\mathcal{F}^2 = \left(\frac{W}{L/2}\right)^2\left[1+\chi^2-2\chi \cos\theta/2\right],
\label{eq:F}
\end{equation}
is a characteristic amplitude of the $\theta$-dependent contribution to the strain arising from displacement in the $yz$-plane.

The elastic energy of the sheet is defined as $\Eelastic = t \int w_{s}dxdy$, where the strain energy density $w_{s} = \int \sigma(\epsilon') d\epsilon'$, and $\sigma$ the longitudinal stress. For a Hookean material under uniaxial loading, $\sigma(\epsilon) = E\epsilon$, hence $w_s = \epsilon^2/2$ after integration. However, we observe experimentally a moderate strain softening of the materials (see Fig.~\ref{fig:model}C), indicating that a Hookean elasticity is not accurate for the entire range of applied stretch . We then use an empirical nonlinear elastic constitutive law $\sigma = E \epsilon(1-\anl \epsilon)$, where $\anl$ is a free parameter adjusted from the experimental torque profile at incipient twisting. After integration of the strain, the strain energy density is $w_{s} = \frac{1}{3}\, E \epsilon^2(\frac{3}{2}-\anl\epsilon)$. Integrating $w_s$ over the sheet domain yields: 
\begin{equation}
 \Eelastic =  \mathcal{E}_0+EtLW \left[ \frac{1}{24}\epsilon_0\left(1-\anl\epsilon_0\right)\mathcal{F}^2 + \frac{1}{640}\left(1-2\anl\epsilon_0\right)\mathcal{F}^4 \right], 
 \label{eq:energy}
\end{equation}
where $\mathcal{E}_0= \frac{1}{3}\,E t L W \epsilon_0^2\left(\frac{3}{2}-\anl \epsilon_0\right)$ is the elastic energy at zero twist. In Eq.~(\ref{eq:energy}), higher order terms in $\mathcal{F}^2$ are neglected. Recalling that $M = d\Eelastic/d\theta$, we then obtain an analytical expression for the torque using the chain rule $M = d\mathcal{F}^2/d\theta\,d\Eelastic/d\mathcal{F}^2$:
\begin{equation}
  M = M_0 \left[\frac{2 \chi W \sin \theta/2}{L}\right]\  \left(\frac{1}{12}\epsilon_0\left(1-\anl\epsilon_0\right) + \frac{1}{160}\left(1-2\anl\epsilon_0\right)\mathcal{F}^2\right),
  \label{eq:Torque_analytic}
\end{equation}
where $M_0 =EW^2t$. Now, assuming small strain ($\epsilon \approx \frac{\Delta L}{L}\ll 1$ and $\anl=0$) and small twist $\theta \ll 1$, we have $\chi \approx 1$, and using Eq.~(\ref{eq:F}), $\mathcal{F}^2 \approx (W\theta /L)^2$. Thus, with Eq.~\ref{eq:Torque_analytic}, we recover the linearly increasing FvK torque at small twist $M_{FvK} = \frac{M_0}{12}\frac{W\theta }{L} \left(  \frac{\Delta L}{L}\right)$, and the linear increase with $\frac{\Delta L}{L}$~\cite{chopin2019extreme}.
To quantify $\anl$, we calculate the ratio of the torsional stiffnesses $M' \equiv \frac{dM}{d\theta}$ and $M'_{FvK} \equiv \frac{dM_{FvK}}{d\theta} = \frac{M_0}{12}\frac{W}{L}\left(\frac{\Delta L}{L}\right)$ at incipient twist. Then we measure $M'$ by a linear fit of the data at $\theta = 0$ for various $\Delta L/L$ and plot the evolution of $M'/M'_{FvK}$ with $\Delta L/L$,  in Fig.~\ref{fig:model}C. We find $M' \approx M'_{FvK}$ for $\Delta L/L < 0.1$, but, for larger strains, the FvK stiffness increasingly overestimates the measured torque. This deviation is a result of strain softening experienced by  cross-linked polymers~\cite{boyce2000constitutive} which we account for by using a nonzero hyperelastic parameter $\anl$. From Eq.~(\ref{eq:Torque_analytic}), our model finds $$\frac{M'}{M'_{FvK}} = \left( 1+\frac{1}{2} \frac{\Delta L}{L}\right)\left(1-\mathcal{A} \epsilon_0\right),$$ which we use to fit our data, and obtain $\anl = 1.6 \pm 0.1$.

Fig.~\ref{fig:model}D shows a comparison of the measured torque as a function of twist with Eq.~(\ref{eq:Torque_analytic}) corresponding to the elasto-geometric torque model valid for various finite strains $\Delta L/L$ incorporating the hyperelastic nature of the material.  Our model can be seen to be in good agreement with the observed torque versus twist angle over a wide range of strains $0-0.3$, and captures the nonmonotonic torsional response, quantitatively, until the onset of self-contact at $\theta \approx 180^{\circ}$. Thus, we find that the nonmonotonicity originates from finite rotation effects in the $yz$-plane, which are essentially captured in Eq.~(\ref{eq:Torque_analytic}) by a sine function derived from $d\mathcal{F}^2/d\theta$. Thus, the geometrical nonlinearities incorporated in our elasto-geometric torque model, (which are missing in the FvK model) are crucial to describe the observed torque. Furthermore, it is noteworthy that the torque peak occurring at increasingly higher $\theta > 90^{\circ}$, with strain is also captured by our model. Considering the limit of very thin sheets for which fold-induced resistance vanishes, $\chi = \cos \frac{\theta}{2}$, we obtain a torque profile $\frac{M_0}{12} \frac{W}{L} \sin \theta \,\epsilon_0(1-\mathcal{A}\epsilon_0)$ which peaks at an angle $\theta = 90^{\circ}$ smaller than in our experiment with finite thickness sheets.  Thus, the quantity $\chi$ in our model encodes the fold-induced resistance to transverse displacement.

Next, we compare the elastic energy obtained in Eq.~(\ref{eq:energy}) as a function of $\theta$ with the measured values in Fig.~(\ref{fig:model}E) after normalizing with the zero twist stretching energy $\Einit$. We find very good agreement showing that the bending energy can be neglected (Fig.~\ref{fig:model}F and Supplementary Materials 8: Comparison Stretching and Bending Energy up to Half-turn Twist). It is noteworthy that, while the bending energy contribution to the  elastic energy is small (note the difference in vertical scale in Fig.~(\ref{fig:model}E and Fig.~\ref{fig:model}F), folding is necessary to achieve a net energy reduction. To demonstrate this, we have plotted the helicoidal elastic energy (dashed line) which corresponds to $\chi=1$ in Fig.~(\ref{fig:model}E) . The energy is observed to grow well above the elastic energy of a folded sheet compared to when $\chi<1$ as in the experiments, showing that folding clearly results in a lower growth in the elastic energy.

\begin{figure}
	\centering
\includegraphics[width=0.65\textwidth]{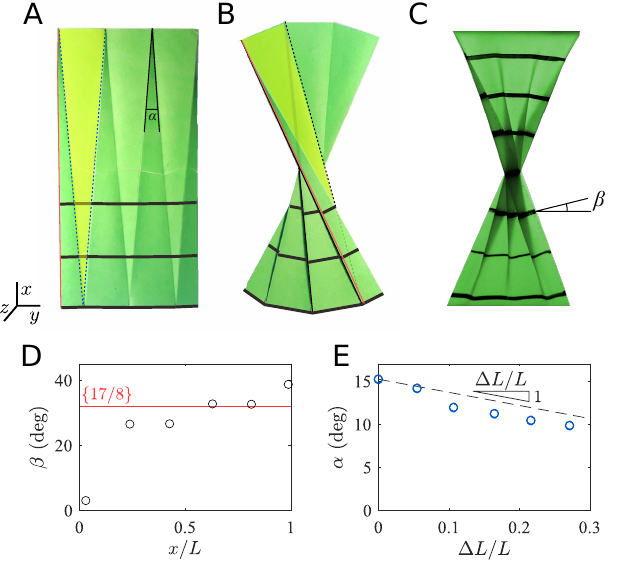}
\caption{
{\bf Half-twisted sheets fold like an origami away from the clamp.} ({\bf A}) Flat sheet with triangular up  and down fold lines. Horizontal black solid line is drawn to indicate the relative displacement. ({\bf B}) Corresponding origami with 6 flat folds. ({\bf C}) Elastic sheet twisted by $\theta = 180^\circ$ shows similar fold structure away from the clamped edges. ({\bf D}) The segment angle $\beta$ as a function of distance across sheet width for elastic sheet and origami. Solid red line indicates the expected segment slope value $\beta=31.8^\circ$ for an origami with the same tip angle $\alpha$. ({\bf E}) The angle $\alpha$ of stretched triangle as a function of applied strain $\Delta L/L$. { A line of slope -1 (dashed line) is shown as a guide to the eye.}}
\label{fig:origami}
\end{figure}

\subsection*{Self-folding and Schl\"afli origami}
To now explain the folded  structure which develops at $\theta = 180^\circ$, we complement our elasto-geometric analysis with origami construction.  Consider an inextensible sheet (Fig.~\ref{fig:origami}A) which can be folded up or down along the dashed lines, resulting in a polygonal spiral origami (Fig.~\ref{fig:origami}B). The apex angle of the isosceles triangular folds is $\alpha$. An image of an elastic sheet with the same aspect ratio is shown in Fig.~(\ref{fig:origami}C), where the thickness of the sheet has been chosen such that it results in the same number of folds as in the origami.  We plot the segment angle $\beta$ from the $y$-axis made by initial horizontal lines in Fig.~(\ref{fig:origami}D). Quantitative agreement is found between the experimental value of $\beta$ away from the clamps, with the expected value (red line) assuming solid body rotation of the triangles, where each fold acts as a hinge. To quantify the role of the stretching on the origami pattern, we measure $\alpha$ from shadowgraph images and find that it decreases with $\Delta L/L$ (Fig.~\ref{fig:origami}E).  This variation follows from the decrease of $\lambda_p$ and the increase in sheet length with stretching, if one assumes $\alpha = \lambda_p/(L+\Delta L)$. Thus, good agreement can be observed between the origami shape and the twisted sheet away from the clamped edges. Further quantitative agreement observed with physical cuts mid-way between the clamps can be found in Supplementary Materials 9: Transect Cut Comparisons.

\begin{figure*}
	\centering
		\includegraphics[width=1\textwidth]{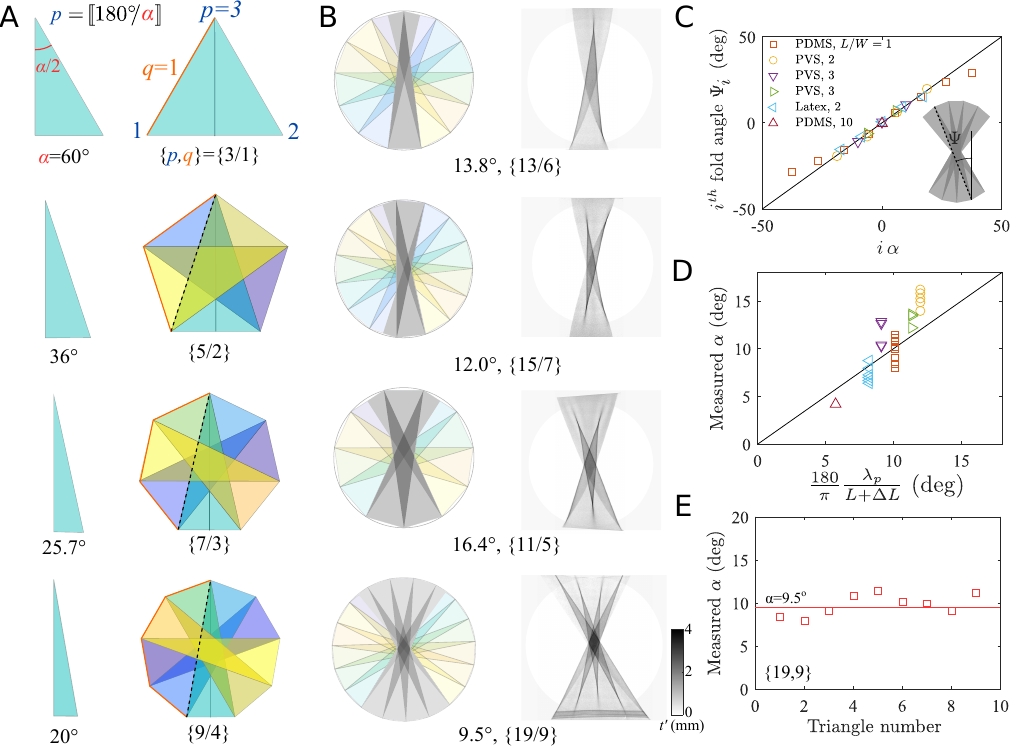}
	\caption{
{\bf Partial Schl\"afli origami explains layered architectures at half-twist.} ({\bf A}) Geometrical forms obtained by increasing the Schl\"afli symbols and number of facets. ({\bf B}) Comparison of the experimental radiogram and Schl\"afli fold origami. Good correspondence is observed in all four cases. ({\bf C}) The angle $\Psi_i$ of the $i^{th}$ fold as a function of the calculated angle $i\,\alpha$ using geometric model is in excellent agreement. ({\bf D}) Comparison of the apex angle $\alpha$ as a function calculated $\alpha$ using various sheets and loading. ({\bf E}) The {apex angle} as a function of {triangle number is essentially constant.}}
	\label{fig:schlaffi}
\end{figure*}

Origami corresponding to spiral accordion folded elastic sheets can be algorithmically generated using $\alpha$ as a parameter. Consider a right angle triangle with height $L$ and angle $\alpha/2$ (Fig.~\ref{fig:schlaffi}A). (This triangle is also the same as that at the far left side of the sheet in Fig.~\ref{fig:origami}A.) To help understand the geometrical transformation leading to a flat folded origami, we preserve the color when reflecting off a right triangle with respect to its height, and change color when reflecting the triangle with respect to its hypotenuse identified as a fold. A flat-fold origami is thus obtained by applying, alternately, these two transformations until reaching a given number of folds $n$. When $180^{\circ}/\alpha$ is odd, these transformations result in a regular flat-folded origami where the triangle bases are the edges of a regular polygon with $p$ vertices (Fig.~\ref{fig:schlaffi}A). This polygon is also the convex envelope of a star-shaped polygon composed by the hypotenuses (dashed black line) connecting vertices separated by $q$ consecutive triangle bases (solid orange line).  These origami can be identified by the so-called Schl\"afli symbols $\{p/q\}$~\cite{coxeter1972regular}, and thus we call them {\it Schl\"afli origami}. By geometric construction, we have $p = [\![180/\alpha]\!]$ and $q = (p-1)/2$, where $[\![\cdot]\!]$ stands for the closest odd integer. When $180/\alpha$ is not odd, the calculated symbols should be understood as the ones yielding the closest Schl\"afli origami for a given $\alpha$. Varying the Schl\"afli symbols (or equivalently reducing the tip angle and increasing the triangle numbers), one can obtain triangle, pentagon, heptagon, and nonagon shaped envelopes.  The thickness of the overlapped regions at the center is given by {$t' = (n+1) t$} and decreases in integer multiples of $t$ toward the edges.

In Fig.~\ref{fig:schlaffi}B we show, in the first column, flat-folded Schl\"afli origami of higher symmetry (in lighter shade) as backgrounds of their incomplete counterparts (in grey) obtained by restricting the number of folds to $n =2\  W/\lambda_p$, keeping $p$ the same. The examples in the 1st, 3rd, and 4th row are symmetric, and the one in the 2nd row is asymmetric. Thus, a  Schl\"afli origami with either symmetry can be generated according to our algorithm, by starting with a right angle triangle reflecting about the hypotenuse and height equal to the calculated number of folds. Partial {Schl\"afli origami} have been denoted with a grey-scale proportional to the number of overlapping domains at the particular location.  They can be compared with radiograms of spiral folded elastic sheets which have the same $L/W$ and $n$ (Fig.~\ref{fig:schlaffi}B, second column). The grey-scale in the radiogram is linearly proportional to absorption encountered along the linear path of the x-rays, and thus can be observed to be consistent with those generated by origami. 

This correspondence is further quantified by measuring, from the radiograms, the fold angle $\psi$ and $\alpha$, after a $180^\circ$ twist (Fig.~\ref{fig:schlaffi}C-E). Notably, $\alpha$ does not vary significantly between the triangles of a given twisted sheet, in accordance with the predictions of the Schl\"afli origami (Fig.~\ref{fig:schlaffi}E). We find an excellent agreement without any fit parameters for all three measures. Thus, the orientation of the folds is given by our model even though it neglects the elastic stretching of the sheet, demonstrating that origami kinematics underpin tensional twist-folding. 

\subsection*{ Secondary instabilities and yarn formation}
\begin{figure*}
	\centering
		\includegraphics[width=.9\textwidth]{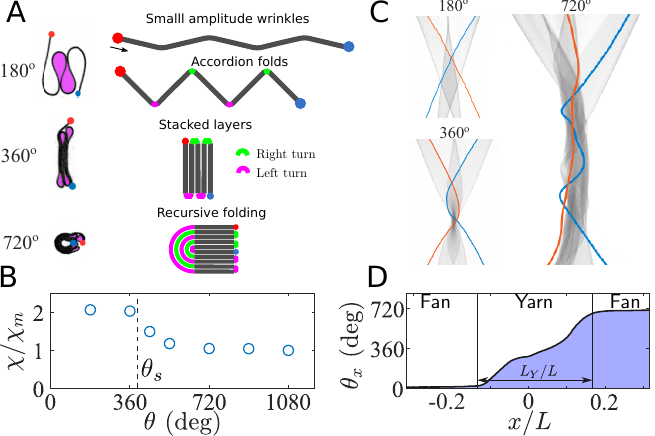}
	\caption{
{\bf Secondary instability and yarn formation.} ({\bf A}) Crosssections at $x/L=0.5$ for twist angle 180, 360, and 720$^\circ$ highlighting edges (red and blue disks) and encapsulated regions (magenta shades). Corresponding schematics illustrating accordion folding and period doubling at the secondary instability. ({\bf B}) Compaction parameter shows a sharp decrease at the secondary instability. ({\bf C}) Fluoroscopy images corresponding to $\theta = 180^\circ, 360^\circ$, and $720^\circ$ with superposed edges winding around each other. ({\bf D}) The orientation angle of the edge $\theta_x$ versus $x/L$. Inset : $\theta_x$ is the angle of the segment joining the two ends of the crosssection. }
	\label{fig:yarn}
\end{figure*}

{We now examine the transformation of the folded sheets into yarns by plotting transects at mid-distance between the clamps for $\theta = 180^\circ$, $360^\circ$, and $720^\circ$ in Fig.~\ref{fig:yarn}A.} The same left and right edges of the sheet are denoted with red and blue markers, respectively. The central helical yarn section undergoes strong compaction 
by $\theta = 360^\circ$, and then folds recursively when a secondary instability occurs at $\theta_s \approx 400^\circ$. Encapsulated regions are highlighted by the magenta 
shade.   The $\chi$ normalized by its minimum value $\chi_m$ is plotted as a function of $\theta$ in Fig.~\ref{fig:yarn}B. We observe that the ratio decreases from 2 to 1 after the secondary instability (denoted by the vertical line), showing a recursive folding of the sheet. 

Then, we represent the features by which the multi-layer yarns form by idealized straight crossectional segments connected by curved joints making right (R) or left (L) turns going from one edge of the sheet to the other by the black arrow in Fig.~\ref{fig:yarn}A. We take the cross-section shown in Fig.~(\ref{fig:yarn}A) as an example. With this convention, the configuration of the cross-section before and after secondary instability is encoded as LRLR (accordion)  and   LLRRLLRRL (folded accordion), respectively. The schematics highlight a period doubling by recursive folding in a way which is qualitatively different from twist-less compressed sheets, where the sequence of turns LRLR, transforms into LRLRLRLR after period doubling~\cite{brau2011multiple,Brau2013}.

Fig.~\ref{fig:yarn}C shows the radiograms at the corresponding $\theta$ where the tracked edges of the sheet are marked in red and blue. { While the folding and helical wrapping yield complex internal structures, the stretched edges are found to wind around each other, similar to the twisting of two filaments into a rope~\cite{Olsen2010}. The crossings between the two edges 
in the projected plane occur in the yarn-like compact region which starts to develop along the longitudinal axis.} 
To quantify the yarn region, we use the orientation of the segment joining the end-points in the $y-z$ plane to obtain a cross-section orientation angle $\theta_x$ (Fig.~\ref{fig:yarn}D, inset and Supplementary Materials: Fig.~S10). We plot $\theta_x$ in Fig.~\ref{fig:yarn}D after the sheet is twisted twice, and find that the twist is localized in a central section $L_{Y}$, where the local twisting rate $\theta_x/L_{Y}$ is approximately 4 times greater than $\theta/L$. 
We use these observations to introduce a geometric model of yarn formation from the accordion folded sheet.

\subsection*{Geometric yarn model}
\begin{figure*}
	\centering
		\includegraphics[width=.6\textwidth]{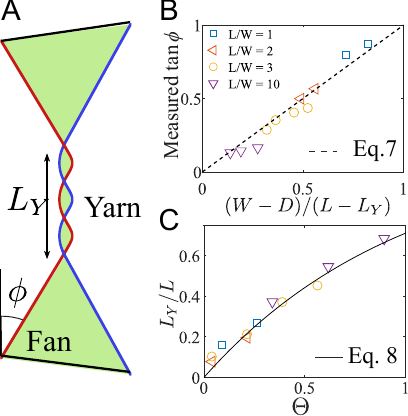}
	\caption{
{\bf Yarn model and comparison.} ({\bf A}) Schematics illustrating the geometric yarn model. ({\bf B}) Measured angle subtended by the fan versus the prediction $\phi=(W-D)/(L-L_Y)$. ({\bf C}) The fraction of the yarn length $L_Y/L$ versus $\Theta$.}
	\label{fig:yarnmodel}
\end{figure*}

In order to model the growth of the yarns, we assume that the sheet can be divided into three sections with a yarn-like structure of length $L_Y$ and two fan-like structures near the clamped edges characterized by fan angle $\phi$, as shown schematically in Fig.~(\ref{fig:yarnmodel}A). This simplification enables us to retain the fundamental role of the twisted sheet-edge in the elasto-geometric torque model, while circumventing the difficulty in calculating strains.

We measure the evolution of $L_Y$ and $\phi$  with twist over various $L/W$, and find that $L_Y$ and $\phi$ increase with $\theta$ (see Supplementary Materials: Fig.~(S11B and C)). The $L_Y$ increases quasi-linearly with twist rate depending on the aspect ratio, with the onset of yarn formation observed to begin after the secondary instability occurs at $\theta = \theta_s$. Using trigonometry, we can express the fan angle as a function of the sheet aspect ratio $W/L$, scaled yarn width $D/W$, and scaled yarn length $L_Y/L$, yielding:
\begin{equation}
\tan \phi = \left(\frac{W}{L}\right) \frac{1-D/W}{1-L_Y/L}.
\label{eq:yarn}
\end{equation}
Before yarn formation $L_Y =0$, $D =0$, and we note from Eq.~(\ref{eq:yarn}) that $\tan \phi(\theta_s) = W/L$, 
where $\phi(\theta_s)$ can be interpreted as the angle that the diagonal makes with the longitudinal axis. This form predicts an overall decrease of $\phi$ with the sheet aspect ratio $L/W$.  We compare $\phi$ obtained using various sheets with Eq.~(\ref{eq:yarn}) in Fig.~\ref{fig:yarnmodel}B, and find very good agreement.

Then, the evolution of the yarn length can be understood from the helical wrapping of the fan edges around a cylindrical core of diameter $D$ that encompasses the crosssection of the compacted material in the yarn region. The fan edges are assumed to be in direct contact with the core with an angle $\phi$ (see Supplementary Materials 11: Yarn shape analysis and Fig.~(S11A)),  thus forming a helix with a local twist rate $\tan \phi/(D/2)$. We further impose the yarn growth rate with twist, $dL_Y/d\theta$ to be set by the local twist rate, yielding $dL_Y/d\theta = \tan \phi/(D/2)$.  Using Eq.~(\ref{eq:yarn}), we find that $L_Y$ is modeled by a linear first-order ODE whose solution is: 
\begin{equation}
\frac{L_Y}{L} = 1-\exp(-\Theta/2),    
\label{eq:growth}
\end{equation}
where $\Theta =\chi\ (\theta-\theta_s)/(1-\chi)$. Considering that there are no adjustable parameters, this growth model is in very good agreement with experimental data shown in Fig.~\ref{fig:yarnmodel}C. 

\section*{Discussion}
Thus, we explain our observation of a remarkably ordered transformation of flat sheets to scrolled multilayered yarns by introducing a series of simplified elasto-geometric models that form our tensional twist-folding framework as summarized in Fig.~\ref{fig:overview}. In the post-wrinkling regime, the oscillating torque profile and associated accordion folding into a multilayered sheet is explained quantitatively by our combined elasto-geometric torque model and Schl\"afli origami kinematics. In the elasto-geometric torque model, the bending energy carried by the folds is negligible compared to the longitudinal stretching energy. Consequently, the accordion shape is approximated by a piece-wise ruled surface with straight crosssections. This backbone shape keeps track of the rotation of the crosssection from one clamp to the other and the lateral compaction which is parametrized by $\chi$ measuring the change in the (projected) mid-crosssection length relative to the sheet width. To calculate the elastic strain energy over the regime of large shape transformation, we use the Green-Lagrange strain incorporating geometric nonlinearities arising from large displacements in three dimensional space. These are fundamental differences from the FvK models used to explain initial wrinkling instability with twist~\cite{Chopin2015}. We find that finite rotation in the $yz$-plane is responsible for the torque saw-tooth profile shown in Fig.~\ref{fig:model}D, while the precise peak location is modulated by the fold-induced transverse stiffness of the sheet captured by Eq.~(\ref{eq:Torque_analytic}).  Important, yet nonessential to capture the oscillation torque profile, we introduce a weakly nonlinear elasticity to include strain softening at larger strain. Notably, the assumption of negligible bending energy is validated experimentally by the excellent agreement between the calculated and measured elastic energy in Fig.~\ref{fig:model}E.

To explain the flat multilayered structure observed after a $180^\circ$ twist, we then introduced a Schl\"afli origami model, where we consider the inextensible sheet limit (represented as in a sheet of paper) in alternating mountains and valley creases. The resulting folds form a sequence of vertically oriented triangles with complementary orientations. Thus, the observed transformed sheet shape is parametrized by the number of creases and the angle of the triangle apex in our model. In the flat folded state, we show that the origami, when twisted by a half-turn, can form regular star-shaped polygons characterized by Schl\"afli symbols. These flat folded Schl\"afli origami accurately predict the observed folded structure when the Schl\"afli  symbols and the vertices number are set using sheet $L$, $W$, and $\lambda_p$. Thus, we find the imprint of the primary wrinkling instability persists far beyond onset. Deviation of the shape from our prediction is observed only near the two clamps, where significant stretching is needed to satisfy the clamped boundary conditions. Nonetheless, it is remarkable that the kinematics obtained by an inextensible origami model is preserved when significant tension is applied, as in the elastic sheets used in our experiments. The key role of sheet stretchability and applied tension in selecting the number of folds, and in organizing the folding into tightly scrolled yarns is thus uncovered by our analysis.  

Further, we developed a geometric yarn model to explain the evolution of the folded sheet after a secondary instability. Based on x-ray tomography analysis, we postulate that the structure can be considered as being composed of a highly twisted yarn region at the center and weakly twisted fan-like regions connected to the two clamps. Our model is based on simplified kinematics where the edges are straight and co-planar in the fan region, but form two helices winding around each other and a compact cylindrical core idealizing the multi-layered yarn.  Our model explains the decrease in size of two fan-like regions upon twist and the linear growth of the yarn length initially, with an exponential slow-down as the yarn-ends approach the clamps given by Eq.~(\ref{eq:growth}). Thus, our investigations document the main stages of the transformation of a flat sheet into multilayered yarns and provide a framework for their analysis.

Our framework enables multifunctional yarns that have been proposed using ultra thin polyethylene sheets, carbon nanotubes, and graphene sheets~\cite{Lima2011,xu2013ultrastrong} towards applications in medical materials and flexible electronics~\cite{Kim2012,Cruz2014,silverberg2014using} to be generated with programmed structure using remote loading. These multilayered architectures and yarns are otherwise difficult to  achieve using in-plane compression or shear alone without further direct manipulations. The PDMS and PVS sheets in our study were used repeatedly because of their hyperelastic nature. This enabled the sheets to be unfolded and reconfigured multiple times during the course of the trials, while recording their shapes and torques under different loading conditions. Thus, our tensional twist-folding strategy can be used to create redeployable functional structures from simple elements with the appropriate choice of materials, an important goal for advanced manufacturing with soft materials~\cite{kwon2014high,Lima2011,silverberg2014using}.

\section*{Materials and Methods}

The material properties of the sheets used in the study can be found in the Supplementary Materials 1: Materials. The shape measurement methods are discussed in Supplementary Materials 2: Optical Imaging, and Supplementary Materials 3: Fluoroscopy and Computed Tomography, respectively.   

The data in the analysis corresponding to the measured sheets, twist, and scans can be found in the main document, and in Ref.~\cite{Chopin2018}.





\section*{Acknowledgments}
We thank Andreea Panaitescu for help with preliminary experiments, and Benjamin Allen and Fabio Lingua for setting up the x-ray system. We thank Madelyn Leembruggen, Benoit Roman, Brian Chang, and Joseph Paulsen for their critical reading of the manuscript and their comments. 

\subsection*{Funding}
U.S. National Science Foundation grants DMR-1508186\\
U.S. National Science Foundation grants DMR-2005090.

\subsection*{Author contributions}
Conceptualization: JC, AK\\
Data curation: AK\\
Funding acquisition: AK\\
Investigation: JC, AK\\
Methodology: JC, AK\\
Software: JC\\
Visualization: JC\\
Writing—review \& editing: JC, AK

\subsection*{Competing interests} The authors declare no competing interests. 

\subsection*{Data and materials availability} All data are available in the main text or the supplementary materials.

\clearpage
\section*{Supplementary materials}
Supporting Information Text 1 to 11\\
Figs. S1 to S11\\
Tables S1 and S2\\
Movie S1\\



\end{document}